# Normalization of zero-inflated data:

# An empirical analysis of a new indicator family

# and its use with altmetrics data[1]

Lutz Bornmann[*] and Robin Haunschild[**]

[*]First author and corresponding author:

Division for Science and Innovation Studies

Administrative Headquarters of the Max Planck Society

Hofgartenstr. 8,

80539 Munich, Germany.

Email: bornmann@gv.mpg.de

[**] Contributing author:

Max Planck Institute for Solid State Research

Heisenbergstr. 1,

70569 Stuttgart, Germany.

Email: R.Haunschild@fkf.mpg.de

---

**Abstract**

Recently, two new indicators (Equalized Mean-based Normalized Proportion Cited, EMNPC; Mean-based Normalized Proportion Cited, MNPC) were proposed which are intended for sparse scientometrics data. The indicators compare the proportion of mentioned papers (e.g. on Facebook) of a unit (e.g., a researcher or institution) with the proportion of mentioned papers in the corresponding fields and publication years (the expected values). In this study, we propose a third indicator (Mantel-Haenszel quotient, MHq) belonging to the same indicator family. The MHq is based on the MH analysis – an established method in statistics for the comparison of proportions. We test (using citations and assessments by peers, i.e. F1000Prime recommendations) if the three indicators can distinguish between different quality levels as defined on the basis of the assessments by peers. Thus, we test their convergent validity. We find that the indicator MHq is able to distinguish between the quality levels in most cases while MNPC and EMNPC are not. Since the MHq is shown in this study to be a valid indicator, we apply it to six types of zero-inflated altmetrics data and test whether different altmetrics sources are related to quality. The results for the various altmetrics demonstrate that the relationship between altmetrics (Wikipedia, Facebook, blogs, and news data) and assessments by peers is not as strong as the relationship between citations and assessments by peers. Actually, the relationship between citations and peer assessments is about two to three times stronger than the association between altmetrics and assessments by peers.

# 1　Introduction

Alternative metrics (altmetrics) have been established as a new fast-moving and dynamic area in scientometrics (Galloway, Pease, & Rauh, 2013). Initially, altmetrics have been proposed as an alternative to traditional bibliometric indicators. Altmetrics are a collection of multiple digital indicators which measure activity related to research papers on social media platforms, in mainstream media, or in policy documents (National Information Standards Organization, 2016; Work, Haustein, Bowman, & Larivière, 2015). Haustein (2016) identified the following seven groups of platforms which are (currently) used for altmetrics: "(a) social networking (e.g., Facebook, ResearchGate), (b) social bookmarking and reference management (e.g., Mendeley, Zotero), (c) social data sharing including sharing of datasets, software code, presentations, figures and videos, etc. (e.g., Figshare, Github), (d) blogging (e.g., ResearchBlogging, Wordpress), (e) microblogging (e.g., Twitter, Weibo), (f) wikis (e.g., Wikipedia), as well as (g) social recommending, rating and reviewing (e.g., Reddit, F1000Prime)" (p. 417).

According to Adie (2014), there are three developments which foster the engagement in altmetrics. (1) Evaluators, funders, or national research assessments are not only interested in research impact inside but also outside of academia (Mohammadi, Thelwall, & Kousha, 2016; Thelwall & Kousha, 2015a). (2) There is a general shift from print to online. In an early study, Bollen, Van de Sompel, and Rodriguez (2008) demonstrated the richness of data from online activities. The data include web citations in digitized scholarly documents and from social media (Wilsdon et al., 2015). (3) The publication of the altmetrics manifesto by Priem, Taraborelli, Groth, and Neylon (2010) gave this new area in scientometrics a name and thus a focal point. Today, many publishers add altmetrics to papers in their collections (e.g., Wiley and Springer) (Thelwall & Kousha, 2015b). Altmetrics are also recommended by Snowball

Metrics (Colledge, 2014) for research evaluation purposes – an initiative publishing global standards for institutional benchmarking in the academic sector (www.snowballmetrics.com).

In recent years, some altmetrics indicators have been proposed which are field- and time-normalized. These indicators were developed because evidences have been published that this data is – similar to bibliometric data – field- and time-dependent (see, e.g., Bornmann, 2014b). Obviously, some fields are more relevant to a broader audience or general public than others (Haustein, Larivière, Thelwall, Amyot, & Peters, 2014). Haunschild and Bornmann (2016) and Bornmann and Haunschild (2016b) introduced the mean discipline normalized reader score (MDNRS) and the mean normalized reader score (MNRS) based on Mendeley data (see also Fairclough & Thelwall, 2015). Bornmann and Haunschild (2016a) propose the Twitter Percentile (TP) – a field- and time-normalized indicator for Twitter data. This indicator was developed against the backdrop of a problem with altmetrics data which is also addressed in this study – the inflation of the data with zero counts. The overview of Work et al. (2015) on studies investigating the coverage of papers on social media platforms show that many platforms have coverages of less than 5% (e.g., blogs or Wikipedia). This result is confirmed by the meta-analysis of Erdt, Nagarajan, Sin, and Theng (2016): their analyses across former empirical studies dealing with the coverage of altmetrics show that about half of the platforms are at or below 5%; except for three (out of eleven) the coverage is below 10%. Common normalization procedures based on averages and percentiles of individual papers are problematic for zero-inflated data sets (Haunschild, Schier, & Bornmann, 2016). Bornmann and Haunschild (2016a) circumvent the problem of zero-inflated Twitter data by including in the calculation of TP only journals with at least 80% of the papers with at least 1 tweet each. However, this procedure leads to the exclusion of many journals.

Recently, Thelwall (2017a, 2017b) proposed another family of field- and time normalized indicators which compares the proportion of mentioned papers (e.g. on Facebook or Wikipedia) of a unit (e.g., a researcher or institution) with the proportion of mentioned papers in the corresponding fields and publication years (the expected values). The family consists of the Equalized Mean-based Normalized Proportion Cited (EMNPC) and the Mean-based Normalized Proportion Cited (MNPC). In this study, we investigate the new indicator family empirically and add a further variant to this family. In statistics, the Mantel-Haenszel (MH) analysis is recommended for pooling the data from multiple 2×2 cross tables based on different subgroups (here: mentioned and not mentioned papers of a unit published in different subject categories and publication years compared with the corresponding reference sets) (Sheskin, 2007). We call the new indicator Mantel-Haenszel quotient (MHq).

In the first step of the empirical analysis, we analyze the convergent validity of the new indicator family by comparing the scores with ratings by peers. We investigate whether the indicators are able to discriminate between different quality levels assigned by peers to publications. Since the convergent validity can only be tested by using citations (which are related to quality), the first empirical part is based on citations. Good performance on the convergent validity test is an important condition for the use of the indicators in altmetrics. For altmetrics, the relationship to quality – as measured by peer assessments – is not clear. Since the first empirical part will show that the MHq is convergent valid, we test the ability of several altmetrics (e.g., Wikipedia and Facebook counts) to discriminate between quality levels. Thus, we investigate whether several altmetrics are related to the quality of publications – measured in terms of peers' assessments.

## 2     Indicators for zero-inflated count data

Whereas the EMNPC and MNPC proposed by Thelwall (2017a) are explained in sections 2.1 and 2.2, the MHq is firstly introduced in section 2.3. The next sections present not only the formulas for the calculation of the three metrics, but also the corresponding 95% confidence intervals (CIs). The CI is a range of possible indicator values: We can be 95% confident that the interval includes the "true" indicator value in the population. With the use of CIs, we assume that we analyse sample data and infer to a larger, inaccessible population (Williams & Bornmann, 2016). According to Claveau (2016), the general argument for using inferential statistics with scientometric data is "that these observations are realizations of an underlying data generating process … The goal is to learn properties of the data generating process. The set of observations to which we have access, although they are all the actual realizations of the process, do not constitute the set of all possible realizations. In consequence, we face the standard situation of having to infer from an accessible set of observations – what is normally called the sample – to a larger, inaccessible one – the population. Inferential statistics are thus pertinent" (p. 1233).

The relationship between 95% CIs and statistical significance (in case of independent proportions) is as follows:

"1. If the 95% CIs on two independent proportions just touch end-to-end, overlap is zero and the p value for testing the null hypothesis of no difference is approximately .01.

2. If there's a gap between the CIs, meaning no overlap, then p<.01.

3. Moderate overlap … of the two CIs implies that p is approximately .05. Less overlap means p<.05.

Moderate overlap is overlap of about half the average length of the overlapping arms" (Cumming, 2012, p. 402).

## 2.1 Equalized Mean-based Normalized Proportion Cited (EMNPC)

Thelwall (2017a, 2017b) introduced the EMNPC as an alternative indicator for zero-inflated count data. It is an advantage of the EMNPC compared to TP that it is not necessary to reduce the publication set under study to that part which has been frequently mentioned (e.g. on Wikipedia). The approach of the EMNPC is to calculate the proportion of papers that are mentioned: suppose that publication set $g$ has $n_{gf}$ papers in the publication year and subject category combination $f$. $s_{gf}$ of the papers are mentioned (e.g. on Wikipedia). $F$ is defined as all publication year and subject category combinations of the papers in the set. The overall proportion of *g*'s papers that are mentioned is the number of mentioned papers ($s_{gf}$) divided by the total number of papers ($n_{gf}$):

$$p_g = \sum_{f \in F} s_{gf} \bigg/ \sum_{f \in F} n_{gf} \qquad (1)$$

However, $p_g$ could lead to misleading results if the publication set *g* includes many papers which are published in fields with many mentioned papers. Thelwall (2017a, 2017b) proposes to avoid the problem by artificially treating *g* as having the same number of papers in each publication year and subject category combination. The author fixes it to the arithmetic average of numbers in each combination, but recommends not including in the analysis combinations of *g* with only a few papers. Thus, the equalized sample proportion of *g*, $\hat{p}$ is the simple average of the proportions in each combination

$$\hat{p}_g = \frac{\sum_{f \in F} \frac{s_{gf}}{n_{gf}}}{[F]} \qquad (2)$$

The corresponding world sample proportion is defined as:

$$\hat{p}_w = \frac{\sum_{f \in F} \frac{s_{wf}}{n_{wf}}}{[F]} \qquad (3)$$

In Eqns. (2) and (3), [F] is the number of subject category and publication year combinations in which the group (in case of Eq. (2)) and the world (in case of Eq. (3)) publishes. Thus, the equalized group sample proportion has the undesirable property that it treats *g* as if the average mentions of its papers did not vary between the subject categories and publication years. The EMNPC for each publication set *g* is the ratio of both equalized sample proportions:

$$\text{EMNPC} = \hat{p}_g / \hat{p}_w \tag{4}$$

CIs for the EMNPC can be calculated as follows (Thelwall, 2017a):

$$EMNPC_L = exp\left(\ln\left(\frac{\hat{p}_g}{\hat{p}_w}\right) - 1.96\sqrt{\frac{(n_g - \hat{p}_g n_g)/(\hat{p}_g n_g)}{n_g} + \frac{(n_w - \hat{p}_w n_w)/(\hat{p}_w n_w)}{n_w}}\right) \tag{5}$$

$$EMNPC_U = exp\left(\ln\left(\frac{\hat{p}_g}{\hat{p}_w}\right) + 1.96\sqrt{\frac{(n_g - \hat{p}_g n_g)/(\hat{p}_g n_g)}{n_g} + \frac{(n_w - \hat{p}_w n_w)/(\hat{p}_w n_w)}{n_w}}\right) \tag{6}$$

Here, $n_g$ is the total sample size of the group and $n_w$ is the total sample size of the world.

In the following, we demonstrate the calculation of the EMNPC by using the small world example in Table 1. This world consists of papers in four subject categories. The papers of two units (publication set A and B) determine the world. For each unit, the numbers of mentioned and not mentioned papers as well as the corresponding proportion of mentioned papers are given. For example, the unit named as publication set A has published 18 mentioned and 13 not mentioned papers in subject category 1. The proportion of the papers mentioned is 0.58.

Table 1. Small world example for the explanation of the Equalized Mean-based Normalized Proportion Cited (EMNPC)

| **World (reference sets)** | Paper is mentioned | Paper is not mentioned | Number of papers | Proportion mentioned | EMNPC with confidence intervals |
|---|---|---|---|---|---|
| Subject category 1 | 44 | 20 | 64 | 0.69 | |
| Subject category 2 | 30 | 16 | 46 | 0.65 | |
| Subject category 3 | 16 | 12 | 28 | 0.57 | |

| | | | | | | |
|---|---|---|---|---|---|---|
| Subject category 4 | 0 | 20 | 20 | 0.00 | | |
| Total | | | 158 | 0.48 | | **1.00 [0.79, 1.26]** |
| | | | | | | |
| **Publication set A** | | | | | | |
| Subject category 1 | 18 | 13 | 31 | 0.58 | | |
| Subject category 2 | 15 | 9 | 24 | 0.63 | | |
| Subject category 3 | 13 | 9 | 22 | 0.59 | | |
| Subject category 4 | 0 | 10 | 10 | 0.00 | | |
| Total | | | 87 | 0.45 | | **0.94 [0.71, 1.25]** |
| | | | | | | |
| **Publication set B** | | | | | | |
| Subject category 1 | 26 | 7 | 33 | 0.79 | | |
| Subject category 2 | 15 | 7 | 22 | 0.68 | | |
| Subject category 3 | 3 | 3 | 6 | 0.50 | | |
| Subject category 4 | 0 | 10 | 10 | 0.00 | | |
| Total | | | 71 | 0.49 | | **1.03 [0.77, 1.37]** |

The EMNPC of the world equals 1 if 0.48 is divided by 0.48. Thus, it is an advantage for the interpretation of the EMNPC that a world average of 1 exists. With EMNPC=1.03 publication set B performed slightly better than the world average and also slightly better than the publication set A with EMNPC=0.94. However, since the CIs of both sets overlap substantially among themselves and with 1 (the world EMNPC), they do not differ statistically significantly from one another and the world average.

**2.2  Mean-based Normalized Proportion Cited (MNPC)**

The second indicator proposed by Thelwall (2017a), MNPC, is calculated as follows: For each paper with at least one mention (e.g., on Wikipedia), the number of mentions is replaced by the reciprocal of the world proportion mentioned for the corresponding subject category and publication year. All other papers with zero mentions remain at zero. Let $p_{gf}=s_{gf}/n_{gf}$ be the proportion of papers mentioned for publication set $g$ in the corresponding subject category and publication year combination $f$ and let $p_{wf}=s_{wf}/n_{wf}$ be the proportion of world's papers cited in the same year and subject category combination $f$. Then

$$r_i = \begin{cases} 0 \text{ if } c_i = 0 \\ 1/p_{wf} \text{ if } c_i > 0, \text{where paper } i \text{ is from year and subject category combination } f \end{cases} \quad (7)$$

Following the calculation of the MNCS (Waltman, van Eck, van Leeuwen, Visser, & van Raan, 2011), the MNPC is defined as:

$$MNPC = \frac{(r_1 + r_2 + \cdots r_{n_g})}{n_g} \quad (8)$$

An approximate CI has been constructed by Thelwall (2016, 2017a) for the MNPC. In the first step, the lower limit $L$ (MNPC$_{fgL}$) and upper limit $U$ (MNPC$_{fgU}$) for group $g$ in subject category and publication year combination $f$ is calculated with:

$$MNPC_{gfL} = \exp\left(\ln\left(\frac{\hat{p}_{gf}}{\hat{p}_{wf}}\right) - 1.96\sqrt{\frac{(n_{gf} - \hat{p}_{gf} n_{gf})/(\hat{p}_{gf} n_{gf})}{n_{gf}} + \frac{(n_{wf} - \hat{p}_{wf} n_{wf})/(\hat{p}_{wf} n_{wf})}{n_{wf}}}\right) \quad (9)$$

$$MNPC_{gfU} = \exp\left(\ln\left(\frac{\hat{p}_{gf}}{\hat{p}_{wf}}\right) + 1.96\sqrt{\frac{(n_{gf} - \hat{p}_{gf} n_{gf})/(\hat{p}_{gf} n_{gf})}{n_{gf}} + \frac{(n_{wf} - \hat{p}_{wf} n_{wf})/(\hat{p}_{wf} n_{wf})}{n_{wf}}}\right) \quad (10)$$

In the second step, the group-specific lower and upper limits are used to calculate the MNPC CIs:

$$MNPC_L = MNPC - \sum_{f \in F} \frac{n_{gf}}{n_g} \left(\frac{p_{gf}}{p_{wf}} - MNPC_{gfL}\right) \quad (11)$$

$$MNPC_U = MNPC + \sum_{f \in F} \frac{n_{gf}}{n_g} \left(MNPC_{gfU} - \frac{p_{gf}}{p_{wf}}\right) \quad (12)$$

The MNPC cannot be calculated, if any of the world proportions are equal to zero. Furthermore, CIs cannot be calculated if any of the group proportions are equal to zero. Thus, Thelwall (2017a) proposed to remove the corresponding subject category publication year combination from the data or to add a continuity correction of 0.5 to the number of mentioned and not mentioned papers in these cases. We prefer the latter (to add 0.5 to the number of papers mentioned and not mentioned, respectively) (see the example in Table 2). This approach is recommended by Plackett (1974) for the calculation of odds ratios.

Table 2 is based on the same small world example for the explanation of the MNPC, which is also used for the explanation of the EMNPC (see Table 1). Using the MNPC formula above, the MNPC for each subject category and the MNPC across the categories have been calculated for the world and both units. As the results in Table 2 point out, publication set B has a slightly higher proportion of mentioned papers (MNPC=1.07) than the world (MNPC=1.00). Correspondingly, the proportion of publication set A (MNPC=0.94) is slightly lower than the world proportion. However, the CIs of both sets overlap substantially among themselves and with 1 (the world MNPC). Thus, they do not differ statistically significantly from one another and the world average.

Table 2. Small world example for the explanation of the Mean Normalized Proportion Cited (MNPC)

| **World (reference sets)** | Paper is mentioned | Paper is not mentioned | Number of papers | Ratio of number of papers and number of mentioned papers | MNPC with confidence interval |
|---|---|---|---|---|---|
| Subject category 1 | 44 | 20 | 64 | 1.45 | 1.00 |
| Subject category 2 | 30 | 16 | 46 | 1.53 | 1.00 |
| Subject category 3 | 16 | 12 | 28 | 1.75 | 1.00 |
| Subject category 4 | 1 | 21 | 22 | 22.00 | 1.00 |
| Total | | | 160 | | **1.00 [0.65, 3.23]** |
| | | | | | |
| **Publication set A** | | | | | |
| Subject category 1 | 18 | 13 | 31 | 1.72 | 0.84 |
| Subject category 2 | 15 | 9 | 24 | 1.60 | 0.96 |
| Subject category 3 | 13 | 9 | 22 | 1.69 | 1.03 |
| Subject category 4 | 0.5 | 10.5 | 11 | 22.00 | 1.00 |
| Total | | | 88 | | **0.94 [0.56, 4.66]** |
| | | | | | |
| **Publication set B** | | | | | |
| Subject category 1 | 26 | 7 | 33 | 1.27 | 1.15 |
| Subject category 2 | 15 | 7 | 22 | 1.47 | 1.05 |
| Subject category 3 | 3 | 3 | 6 | 2.00 | 0.88 |
| Subject category 4 | 0.5 | 10.5 | 11 | 22.00 | 1.00 |

| Total | | | 72 | **1.07 [0.67, 5.51]** |

### 2.3 Mantel-Haenszel quotient (MHq)

For pooling the data from multiple 2×2 cross tables based on different subgroups (which are part of a larger population), the most commonly used and recommended method is the MH analysis (Hollander & Wolfe, 1999; Mantel & Haenszel, 1959; Sheskin, 2007). According to Fleiss, Levin, and Paik (2003), the method "permits one to estimate the assumed common odds ratio and to test whether the overall degree of association is significant. Curiously, it is not the odds ratio itself but another measure of association that directly underlies the test for overall association … The fact that the methods use simple, closed-form formulas has much to recommend it" (p. 250). Radhakrishna (1965) demonstrate that the MH approach is formally and empirically valid against the background of clinical trial.

The MH analysis results in a summary odds ratio for multiple 2×2 cross tables which we call MHq. For the impact comparison of units in science with reference sets, the 2×2 cross tables (which are pooled) consist of the number of papers mentioned and not mentioned in subject category and publication year combinations $f$. Thus, in the 2×2 subject-specific cross table with the cells $a_f$, $b_f$, $c_f$, and $d_f$ (see Table 2), $a_f$ is the number of mentioned papers published by unit $g$ in subject category and publication year $f$, $b_f$ is the number of not mentioned papers published by unit $g$ in subject category and publication year $f$, $c_f$ is the number of mentioned papers in subject category and publication year $f$, $d_f$ is the number of not mentioned papers published in subject category and publication year $f$. Note that the papers of group $g$ are also part of the papers in the world. In section 4.2, we discuss the possibility that group $g$ is not part of the world. In bibliometrics, however, it is usual that the world consists of all papers published in subject category and publication year $f$.

Table 3. 2 x 2 subject-specific cross table

|         | Number of mentioned papers | Number of not mentioned papers |
|---------|----------------------------|--------------------------------|
| Group g | $a_f$                      | $b_f$                          |
| World   | $c_f$                      | $d_f$                          |

We start by defining some dummy variables for the MH analysis:

$$R_f = \frac{a_f d_f}{n_f} \text{ and } R = \sum_{f=1}^{F} R_f, \tag{13}$$

$$S_f = \frac{b_f c_f}{n_f} \text{ and } S = \sum_{f=1}^{F} S_f, \tag{14}$$

$$P_f = \frac{a_f + d_f}{n_f} \text{ and } Q_f = 1 - P_f \tag{15}$$

Where $n_f = a_f + b_f + c_f + d_f$

MHq is simply:

$$\text{MHq} = \frac{R}{S} \tag{16}$$

The MHq is calculated with the group *g* included in the world. We refer to the indicator as MHq' in section 4.2 when it is calculated with the world excluding the group ($c_f'$ = $c_f$ - $a_f$ and $d_f'$ = $d_f$ - $b_f$). The CIs for MHq are calculated following Fleiss et al. (2003). The variance of ln MHq is estimated by:

$$\widehat{Var}(\ln MHq) = \frac{1}{2}\left\{\frac{\sum_{f=1}^{F} P_f R_f}{R^2} + \frac{\sum_{f=1}^{F}(P_f S_f + Q_f R_f)}{RS} + \frac{\sum_{f=1}^{F} Q_f S_f}{S^2}\right\} \tag{17}$$

The CI for the MHq can be constructed with

$$MHq_L = \exp\left[\ln(MHq) - 1.96\sqrt{\widehat{Var}[\ln(MHq)]}\right] \quad (18)$$
$$MHq_U = \exp\left[\ln(MHq) + 1.96\sqrt{\widehat{Var}[\ln(MHq)]}\right] \quad (19)$$

Table 4. Small world example for the Mantel-Haenszel quotient (MHq)

| World (reference sets) | Paper is mentioned | Paper is not mentioned | Number of papers | MHq |
|---|---|---|---|---|
| Subject category 1 | 44 | 20 | 64 | |
| Subject category 2 | 30 | 16 | 46 | |
| Subject category 3 | 16 | 12 | 28 | |
| Subject category 4 | 0 | 20 | 20 | |
| Total | | | | **1.00 [0.61, 1.64]** |
| | | | | |
| **Publication set A** | | | | |
| Subject category 1 | 18 | 13 | 31 | |
| Subject category 2 | 15 | 9 | 24 | |
| Subject category 3 | 13 | 9 | 22 | |
| Subject category 4 | 0 | 10 | 10 | |
| Total | | | | **0.81 [0.46, 1.44]** |
| | | | | |
| **Publication set B** | | | | |
| Subject category 1 | 26 | 7 | 33 | |
| Subject category 2 | 15 | 7 | 22 | |
| Subject category 3 | 3 | 3 | 6 | |
| Subject category 4 | 0 | 10 | 10 | |
| Total | | | | **1.30 [0.66, 2.53]** |

We used the same data as in Table 1 and Table 2 to produce a small world example for explaining the MHq. This example is presented in Table 4. The MHq in the table can be interpreted as follows: the chances of the papers in publication set A of being mentioned (e.g. on Wikipedia) are 0.81 times as large as the world's papers chances. The chances of the papers in publication set B of being mentioned are 1.3 times greater than the world's papers chances. An MHq value equal to 1.0 indicates that there is no difference between the chances of the publication set (A or B) and the reference sets (i.e., the world) of being mentioned. An MHq value less than 1.0 indicates lower chances for the publications in the set of being mentioned compared with the reference sets. Expressed as percentages, the difference between publication set B and the world is

$$100 * (1.3 - 1.0) = 30\% \qquad (20)$$

Thus, the publications in set B have 30% higher chances for being mentioned than the world's publications. Equivalently, publication set B has had 1.3 (1.3/1) times the impact of publication set A. We recommend the calculation of percentages especially in those cases in which the MHq is smaller than 2. The proper interpretation of percentages becomes difficult with higher values.

Similar to the EMNPC and MNPC, it is an advantage of the MHq that the world average has a value of 1. It is a further advantage of the MHq that the result can be expressed as a percentage which is relative to the world average.

We added also CIs to the MHq in Table 4. Since the CIs of both publication sets (A and B) overlap substantially among themselves and with 1.0 (the world MHq), they do not differ statistically significantly from one another and the world average.

## 3  Data sets used

We used the papers of the Web of Science (WoS) from our in-house database – derived from the Science Citation Index Expanded (SCI-E), Social Sciences Citation Index (SSCI), and Arts and Humanities Citation Index (AHCI) provided by Clarivate Analytics (formerly the IP and Science business of Thomson Reuters). All papers of the document type "article" with DOI published between 2010 and 2013 were included to study the indicators. Citations with a three-year citation window are retrieved from our in-house database. We decided to use a fixed citation window of three years: (1) three years are recommended as the minimum citations window for reliable citation analyses (Glänzel & Schoepflin, 1995). (2) Longer citation windows would lead to more papers with at least one citation, i.e. even less sparse data. For field classification, we used the overlapping WoS subject categories (Rons, 2012, 2014).

We matched the publication data with peers' recommendations from F1000Prime. F1000Prime is a post-publication peer review system of papers from mainly medical and biological journals (Bornmann, 2014b, 2015b). Papers are selected by a peer-nominated global "Faculty" of leading scientists and clinicians who then rate the papers and explain their importance. Thus, only a restricted set of papers from the papers in these disciplines covered is reviewed, and most of the papers are actually not. At present, the Faculty numbers more than 5,000 experts worldwide. Faculty members can choose and evaluate any paper that interests them. Although many papers published in popular and high-profile journals (e.g. *Nature*, *New England Journal of Medicine*, *Science*) are rated, 85% of the papers selected are published in specialized or less well-known journals (Wouters & Costas, 2012). The papers are rated by the Faculty members as "Recommended," "Must read" or "Exceptional" which is equivalent to recommendation scores (RSs) of 1, 2, or 3, respectively.

Papers can be recommended multiple times. Therefore, we calculated an average RS, referred to as $\overline{\text{FFa}}$:

$$\overline{\text{FFa}} = \frac{1}{i_{\max}} \sum_{i}^{i_{\max}} \text{FFa}_i \qquad (21)$$

The papers are categorized depending on their $\overline{\text{FFa}}$ value:
- Not recommended papers (Q0): $\overline{\text{FFa}} = 0$. Q0 includes the papers which, even though they may be cited or mentioned, do not have any F1000Prime recommendation.
- Recommended papers with a rather low average score (Q1): $0 < \overline{\text{FFa}} \leq 1.0$
- Recommended papers with a rather high average score (Q2): $\overline{\text{FFa}} > 1.0$

Table 5. Number of papers and proportion of not cited or not mentioned papers, respectively, broken down by data source, publication year and $\overline{FFa}$ groups

| Year | $\overline{FFa}$ | Citations | | Twitter | | Wikipedia | | Facebook | | Policy documents | | Blogs | | News |
|---|---|---|---|---|---|---|---|---|---|---|---|---|---|---|
| | | Number of papers | Proportion not cited | Number of papers | Proportion not mentioned | Number of papers | Proportion not mentioned | Number of papers | Proportion not mentioned | Number of papers | Proportion not mentioned | Number of papers | Proportion not mentioned | Number of papers |
| 2010 | Q0 | 628,862 | 10.36 | 627,082 | 95.63 | 622,505 | 97.95 | 615,467 | 98.52 | 476,612 | 99.40 | 609,015 | 97.68 | 575,740 |
| 2010 | Q1 | 6576 | 0.84 | 6630 | 86.50 | 6559 | 93.15 | 6528 | 95.44 | 5870 | 98.57 | 6479 | 88.29 | 6266 |
| 2010 | Q2 | 4368 | 0.43 | 4413 | 76.21 | 4384 | 86.27 | 4361 | 91.45 | 3982 | 98.32 | 4355 | 76.51 | 4224 |
| 2011 | Q0 | 681,749 | 10.61 | 683,815 | 87.99 | 671,612 | 98.23 | 676,824 | 97.10 | 478,021 | 99.42 | 662,518 | 97.25 | 643,744 |
| 2011 | Q1 | 6324 | 1.12 | 6439 | 69.13 | 6378 | 93.73 | 6393 | 89.86 | 5625 | 98.93 | 6296 | 88.06 | 6149 |
| 2011 | Q2 | 4418 | 0.68 | 4494 | 51.91 | 4476 | 85.50 | 4491 | 79.69 | 4005 | 98.18 | 4450 | 74.38 | 4412 |
| 2012 | Q0 | 733,813 | 10.41 | 737,074 | 72.47 | 724,701 | 98.50 | 734,471 | 93.60 | 538,791 | 99.50 | 720,941 | 96.76 | 706,317 |
| 2012 | Q1 | 5826 | 1.08 | 5974 | 38.47 | 5896 | 94.84 | 5958 | 79.05 | 5227 | 98.68 | 5897 | 87.64 | 5797 |
| 2012 | Q2 | 5042 | 0.46 | 5176 | 23.59 | 5135 | 89.27 | 5171 | 63.80 | 4585 | 98.56 | 5148 | 74.98 | 5098 |
| 2013 | Q0 | 785,961 | 10.84 | 788,706 | 68.12 | 770,850 | 98.76 | 787,195 | 91.22 | 511,479 | 99.58 | 779,485 | 96.45 | 777,566 |
| 2013 | Q1 | 4176 | 1.39 | 4254 | 31.29 | 4192 | 96.61 | 4250 | 71.20 | 3566 | 99.19 | 4200 | 86.98 | 4198 |
| 2013 | Q2 | 6361 | 0.50 | 6512 | 21.10 | 6477 | 91.85 | 6514 | 60.07 | 5564 | 98.85 | 6446 | 73.74 | 6465 |
| Total | | 2,873,476 | 10.42 | 2,880,569 | 79.67 | 2,833,165 | 98.28 | 2,857,623 | 94.61 | 2,043,327 | 99.46 | 2,815,230 | 96.76 | 2,745,976 |

We only included fields where a paper with an F1000Prime recommendation is assigned to, following Waltman and Costas (2014). In order to avoid statistical and numerical problems, we include only fields in the analysis where (1) at least 10 papers are assigned to and (2) the number of cited/mentioned and not cited/not mentioned papers is non-zero. Table 5 shows the number of papers which are included in the analysis and proportion of not cited or not mentioned papers, respectively, broken down by publication year, data source, and $\overline{FFa}$ group.

The results demonstrate that Wikipedia, Facebook, policy documents, blogs, and news have more than 90% of papers with no mentions. This proportion is reduced to around 80% for Twitter; for citation impact, the number of non-cited papers is only around 10%. The results for the different metrics point out that zero-inflation affects citation counts to a much lesser degree than it affects altmetrics. This limitation cannot be completely avoided in this study. Zero-inflated citation data could be provoked by reducing the citation window. A minimum citation window of three years is, however, necessary to allow a meaningful comparison between citation counts and assessments by peers. We expect that impact measurements based on less than three years do not allow the use of citation counts as proxies of quality.

Altmetrics data were added from a locally maintained database with data shared with us by the company Altmetric on June 04, 2016.

In recent years, many studies on altmetrics have calculated the correlation between citations and altmetrics. These studies were interested in the question whether altmetrics measure the same kind of impact as citations (i.e. impact on academia) or another kind of impact (e.g., beyond academia, see Bornmann, 2014a). The idea behind these studies is that "any source measuring any type of scientific impact ought to correlate with some recognized measure of scientific impact, and WoS citations are the main metric used for this purpose" (Li, Thelwall, & Giustini, 2012, p. 465). Bornmann (2015a) conducted a meta-analysis of studies which have investigated correlations between the following three altmetrics and citations: microblogging (Twitter), online reference managers (Mendeley and CiteULike), and blogging. The corresponding correlation coefficients for the meta-analysis were taken from a range of different studies. The meta-analysis calculates a pooled coefficient which allows a generalized statement on the correlation between a specific kind of altmetrics and citations. The results are as follows: "the correlation with traditional citations for micro-blogging counts is negligible (pooled $r = 0.003$), for blog counts it is small (pooled $r = 0.12$) and for bookmark counts from online reference managers, medium to large (CiteULike pooled $r = 0.23$; Mendeley pooled $r = 0.51$)" (p. 1123). Thus, Twitter data seems to have nearly no relationship to citations.

In this study, we investigate six altmetrics and their relationship to peers' assessments:

(1) The most popular microblogging platform is **Twitter** (www.twitter.com), which was founded in 2006. Until recently, users tweeted of up to 140 characters to their followers; up to 280 characters are possible now. Tweets can contain links or references to scientific publications.

(2) **Wikipedia** (www.wikipedia.com) is a multilingual, web-based, and free encyclopedia with openly editable content (Mas-Bleda & Thelwall, 2016). Contributors to Wikipedia often include references to academic publications to support their statements.

(3) A relatively new form of altmetrics is mentions of publications in **policy-related documents**. Recently, Altmetric (www.altmetric.com) has developed a text-mining solution to discover mentions of publications in policy documents and has started to make this data available (Bornmann, Haunschild, & Marx, 2016; Haunschild & Bornmann, 2017).

(4) One of the oldest social media platforms are **blogs** which are online narratives (Bik & Goldstein, 2013). Scholarly bloggers frequently write blogs of very different lengths about papers published in peer reviewed journals (Shema, 2014). These blogs allow extended informal discussions about research (Shema, Bar-Ilan, & Thelwall, 2012), which is referenced in blogs in a formal or informal way.

(5) **Facebook** is one of the most widely used social media and social networking platforms (Bik & Goldstein, 2013). Facebook users can share information on publications with others.

(6) **News** attention is linked to publications via direct links or unique identifiers, such as DOIs (Priem, 2014). Mentions of scientific works in news publishers (e.g. by the New York Times) are counted (see https://www.altmetric.com).

The quantitative literature analysis of Erdt et al. (2016) shows that Twitter (24%) has the highest coverage for papers, followed by Facebook (8%). Wikipedia (3%), blogs (4%), and news (2%) which have very low coverages. According to the results of Haunschild and Bornmann (2017), policy-related documents have with 0.5% an even lower coverage of papers. We did not include Mendeley counts in this study (Mendeley is a popular online reference manager), because Mendeley is not a zero-inflated data source; it has the best coverage among altmetrics data. Erdt et al. (2016) found a pooled coverage across 15 different studies of 59%.

# 4 Results

## 4.1 Convergent validity of the new indicator family

The comparison of indicators with peer evaluation has been widely acknowledged as a way of investigating the convergent validity of metrics (Garfield, 1979; Kreiman & Maunsell, 2011). Convergent validity is the degree to which two measurements of constructs (here: two proxies of scientific quality), which should be theoretically related, are empirically related. Thelwall (2017b) justifies this approach as follows: "If indicators tend to give scores that agree to a large extent with human judgements then it would be reasonable to replace human judgements with them when a decision is not important enough to justify the time necessary for experts to read the articles in question. Indicators can be useful when the value of an assessment is not great enough to justify the time needed by experts to make human judgements" (p. 4). Several publications investigating the relationship between citations and Research Excellence Framework (REF) outcomes report considerable relationships in several subjects such as biological science, psychology, and clinical sciences (Butler & McAllister, 2011; Mahdi, d'Este, & Neely, 2008; McKay, 2012; Smith & Eysenck, 2002; Wouters et al., 2015). Similar results were found for the Italian research assessment exercise: "The correlation strength between peer assessment and bibliometric indicators is statistically significant, although not perfect. Moreover, the strength of the association varies across disciplines, and it also depends on the discipline internal coverage of the used bibliometric database" (Franceschet & Costantini, 2011, p. 284). The overview of Bornmann (2011) shows that a higher citation impact of papers is to be expected with better recommendations from peers.

In recent years, the correlation between the F1000Prime RSs and citation impact scores has already been explored in several studies. The results of the regression model of Bornmann (2015b) demonstrate that about 40% of publications with RS=1 belong to the 10% most frequently cited papers, compared with about 60% of publications with RS=2 and about 73% of publications with RS=3. Waltman and Costas (2014) found "a clear correlation between F1000 recommendations and citations" (p. 433). The meta-analysis of Bornmann (2015b) points out a pooled $r = 0.246$ for the correlation between RSs and citations (based on six correlation coefficients from four studies). The previous results on F1000Prime allow the prognosis, therefore, that citation-based indicators differentiate more or less clearly between the three RSs. In other words, the validity of new indicators can be questioned if they fail to properly differentiate between the three $\overline{FFa}$ groups.

Against this backdrop, we investigate in the current study the ability of the three indicators for zero-inflated count data to differentiate between the $\overline{FFa}$ groups (Q0, Q1, and Q2). We start with the newly introduced MHq indicator. Figure 1 shows the MHqs with 95% CIs for the three $\overline{FFa}$ groups across four publication years. It is clearly visible that the MHqs are very different for the groups. This is an indication for the convergent validity of the MHq: The mean MHq across the years is close to 1 for Q0. The mean MHq for Q1 is about eight times and that for Q2 is about 15 times higher than the mean MHq for Q0. It seems that the MHq indicator significantly separates between the different $\overline{FFa}$ quality levels.

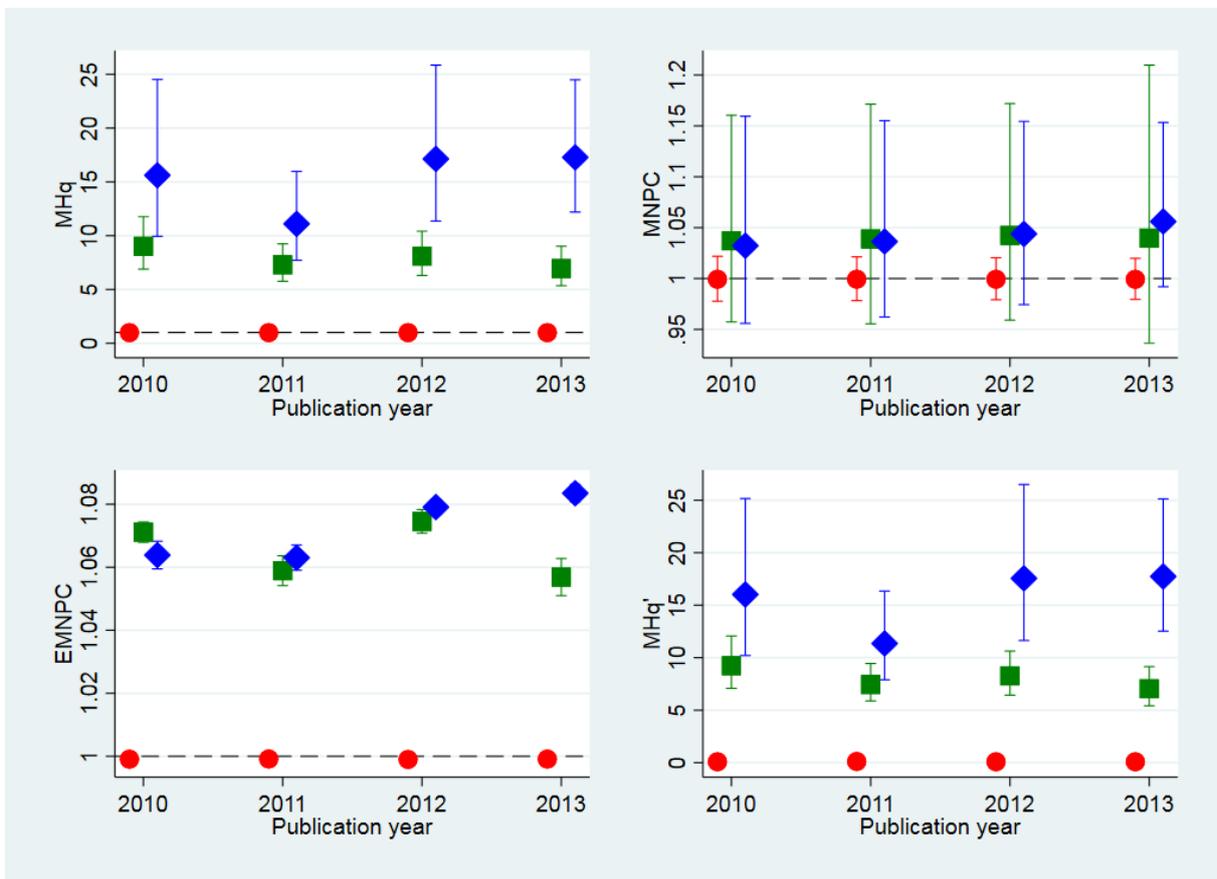

Figure 1. MHq, MNPC, EMNPC, and MHq' with CIs for three $\overline{FFa}$ groups (Q0=red circles, Q1=green squares, and Q2=blue diamonds) and four publications years. The horizontal line with value 1 is the worldwide average.

However, let us take a closer look at the MHq differences between the $\overline{FFa}$ groups on the basis of their CIs following the rules of Cumming (2012) and Cumming and Finch (2005).

If there is a gap between two CIs in the figure, then the difference is statistically significant ($p<.01$). This is the case for the years 2012 and 2013. Here, the indicator differentiates clearly and statistically significantly between the $\overline{FFa}$ groups ($p<.01$). In 2010 and 2011, there is also a statistically significantly difference between Q0 and the other two groups. However, the CIs for Q1 and Q2 overlap in 2010 and 2011. If the overlap between the CIs is less than 50%, then the difference is statistically significant on the $p<.05$ level. This rule is reasonably accurate, however, when the two margins of error (length of one arm of a CI) do not differ by more than a factor of 2. The calculation of the overlaps yields an overlap of 43% in 2010 and 57% in 2011. Thus, the difference between the MHqs is statistically not significant in 2011 ($p>.05$). Although the difference is statistically significant in 2010 ($p<.05$), we cannot assume that the rule works accurately, because the two margins of error differ by a factor of 2.1.

Figure 1 also shows the comparisons between different $\overline{FFa}$ groups for the MNPC and EMNPC – the two indicators proposed by Thelwall (2017a). For both indicators, it is striking that all values in the graphs are very close to 1 – independent of the $\overline{FFa}$ group. This is very different to the MHq, for which the values significantly differ from 1 for the two groups with recommendations (Q1 and Q2). This can be interpreted as a first sign that the MNPC and EMNPC do not differentiate between the quality levels in terms of $\overline{FFa}$ groups. The CIs for the MNPCs in Figure 1 further reveal that the differences between the RSs are not statistically significant. There are clear overlaps for all CIs. The results for the EMNPC in the figure are very heterogeneous. In 2010, the mean value of Q2 is lower than the mean value of Q1. In 2013, the situation is reversed and in the expected direction then. In 2011 and 2012, the mean values are also in the expected direction, but there is a substantial overlap of the CIs (52% in 2012). According to the rules of Cumming (2012) and Cumming and Finch (2005), the differences between the CIs in in both years are statistically not significant.

The world in the MHq analysis which is the reference set can be defined by considering all papers in a certain publication year and field. As we have already pointed out in previous sections, it would also be possible to define the world by excluding the group's papers from the world during calculation of the group's MHq. If the group's papers are included, the group and world are dependent sets of papers. One usually tries to avoid these dependencies in statistics; many models assume that the empirical data are independent. We call this MHq variant MHq', which is also included in Figure 1. The comparison of the results between MHq and MHq' in the figure shows very similar values for Q1 and Q2 as well as for their CIs. Only the values for Q0 considerably differ: MHq' is no longer close to 1 – the worldwide average – but close to 0, because Q0 is compared with a reference set consisting of Q1 and Q2. Thus, it performs significantly worse.

In principle, MHq and MHq' can be calculated for assessing publication sets. The use of the variants depends on the underlying research question. MHq should be calculated, if a group is compared with a reference set (the world). If a group is compared, however, with the rest in the corresponding world, MHq' can be calculated instead. In the calculations of MHq values for altmetrics (see section 4.2), we abstain from excluding the group's papers from the world, because of the following two reasons: (1) Field-normalization in bibliometrics always includes the group's papers in the world. We are not aware of any approach of field-normalization, which exclude the group's papers. (2) The exclusion of the group's papers would mean that the comparison with a reference set would change from a comparison of a group with the world to a comparison of two groups. The world would no longer exist in the calculation of MHq'.

### 4.2   Relationship between altmetrics and the quality of papers

In section 4.1, we demonstrate by using citation data that the MHq is convergent valid, i.e. the indicator is able to discriminate between different quality levels – as defined by peers' assessments. The MHq belongs to the family of indicators for zero-inflated count data and is thus especially designed for altmetrics data. In this section, we use again F1000Prime data to investigate several popular altmetrics whether they are able to discriminate between quality levels – as defined by peers' assessments. Until now, the meaning of altmetrics is one of the most important unanswered questions in scientometric research (Committee for Scientific and Technology Policy, 2014; Haustein et al., 2014; Zahedi, Costas, & Wouters, 2014). We contribute to the open question as we ask whether at least the relationship of research quality and altmetrics can be established at the level of our data set.

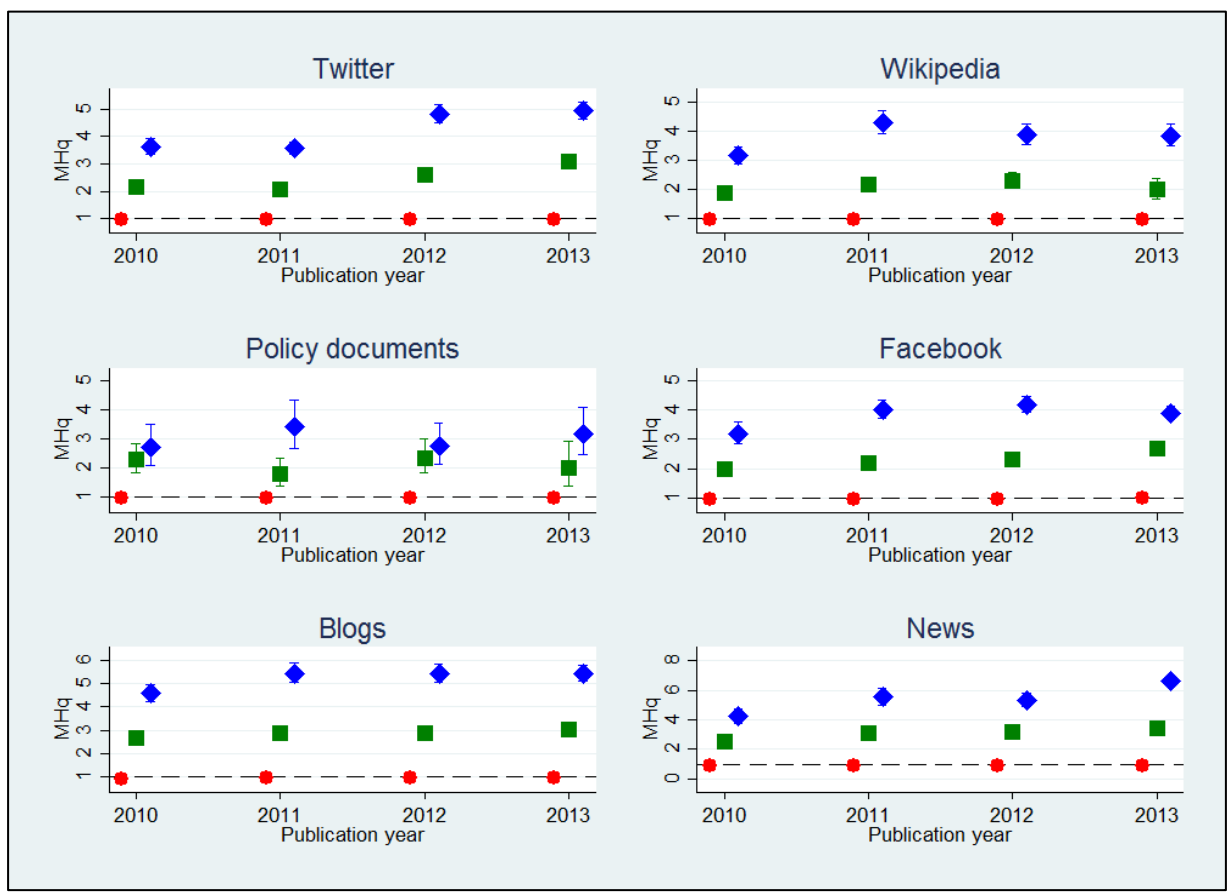

Figure 2. MHqs of six altmetrics with CIs for three $\overline{FFa}$ groups (Q0=red circles, Q1=green squares, and Q2=blue diamonds) and four publications years. The horizontal line with MHq=1 is the worldwide average.

Figure 2 shows the MHqs of six altmetrics with CIs for the three $\overline{FFa}$ groups and four publication years. Since the MHqs for citations are Q0=0.99, Q1=7.84, and Q2=15.28 (means across all years, see section 4.1), the MHqs for the altmetrics are on a significantly lower level, if the $\overline{FFa}$ group is Q1 or Q2. Publications in Q1 and Q2 have about 8 and 15 times higher chances for being cited than the world's populations. These chances are significantly reduced for altmetrics: Papers in Q1 have between about twice (Wikipedia) and about three times (news) the chance as the world's papers for being mentioned; for papers in Q2 these chances are between three times (policy documents) and 5.4 times (news). Thus, the relationship between altmetrics and assessments by peers is not as strong as the relationship between citations and assessments by peers. Actually, the relationship between citations and peer assessments is about two to three times stronger than the association between altmetrics and assessments by peers.

In the comparison of the six altmetrics in Figure 2, it is noticeable that the MHqs are on a somewhat similar level. Thus, the relationship to quality seems to be similarly given. The results further reveal that the differences of MHqs between the $\overline{FFa}$ groups of all altmetrics except policy documents are statistically significant: the CIs do not overlap. The statistics for MHqs for policy-related documents are different, as the CIs of Q1 and Q2 overlap in three of four years substantially (the overlap between Q1 and Q2 is 59% in 2013). According to the rules of Cumming (2012) and Cumming and Finch (2005), the differences between the CIs in the three years are not statistically significant.

## 5  Discussion

The objective of our study is on developing indicators for sparse data, i.e., zero-inflated count data. According to Neylon (2014), much of the altmetrics data we have is sparse. An indicator with many zero values is unlikely to be informative about a scientific unit (e.g. a researcher or institution) in the first place (Thelwall, Kousha, Dinsmore, & Dolby, 2016). Thus, Thelwall (2017a, 2017b) proposed the new family of field- and time-normalized indicators which are especially designed for the use with sparse data. The family consists of the EMNPC and MNPC indicators. Basically, the indicators compare the proportion of mentioned papers of a unit with the proportion of mentioned papers of the world in the corresponding fields and publication years (the expected values).

The indicators of the new family differ from most of the other indicators which have been proposed in bibliometrics and altmetrics hitherto. The other indicators are calculated for single publications and the user of the indicators can aggregate the indicator values (by averaging, summing, etc.). The indicators of the new family are not calculated for single publications, but field- and time-specific publication sets of groups (e.g., single researchers or institutes). Thus, these indicators cannot be used as flexible as the other bibliometric and altmetric indicators. However, we think that it will never be possible to develop reliable indicators with values for single publications for zero-inflated count data.

In this study, we analyze the new indicator family empirically and add a further indicator variant – the MHq. Before the indicators can be used with altmetrics data, they have to be validated and this can only be done on the basis of citation data. Citation data allows formulating predictions which can empirically be validated with the new indicators. Thus, we test with citation data whether the indicators are able to differentiate validly between several quality levels – as defined by F1000 RSs ($\overline{FFa}$). Thus, we compare the indicator values with ratings by peers: Are the indicators able to discriminate between different quality levels which have been assigned by peers to publications?

For the study, citations with a three-year citation window are retrieved from our in-house database as a compromise between having a significant correlation with quality (in the sense of post-publication peer assessments) and having a data set with rather many non-cited papers. Longer citation windows lead to more cited papers and higher correlations with peer assessments. The results for the EMNPC and MNPC show that they cannot discriminate between the different quality levels. The scores for all quality levels are close to 1 (the worldwide average) and the CIs substantially overlap in many comparisons. Thus, the results point out that both the EMNPC and MNPC lack convergent validity. In this study, we further introduced the MHq to the new indicator family which is based on the MH analysis – an established method for pooling the data from multiple 2×2 cross tables based on different subgroups. Since the MHq was able to discriminate empirically between the different quality levels – in most of the cases statistically significant – the convergent validity of the new variant seems to be established.

With MHq' we proposed a variant of the MHq indicator, in which the group's papers are excluded from the world. This variant can be used if the group's paper are compared with the rest in the world. Since in bibliometrics the focus is usually on comparing a group with a reference set, in most of the applications the MHq is the correct choice.

Since the MHq has shown in this study to be a valid indicator (on the basis of F1000Prime recommendations), we applied it to six types of zero-inflated altmetrics data and tested whether different altmetrics sources are related to quality. A substantial relationship to quality is a prerequisite, if the indicator is intended to be used in research evaluation. This study follows calls from other researchers for clarifying the meaning of altmetrics (Priem, 2014; Sugimoto, 2016; Taylor, 2013). "Since altmetrics is still in its infancy, at the moment, we don't yet have a clear definition of the possible meanings of altmetric scores" (Zahedi et al., 2014, p. 1510). According to Thelwall and Kousha (2015b), it is the task of scientometricians to demonstrate that "any given social media metric can be used as an impact indicator" (p. 609). The study of convergent validity is of central importance in this strive for the meaning of altmetrics (Zahedi et al., 2014).

The investigation of the relationship between altmetrics and assessments by peers in this study demonstrates that the relationship between altmetrics and peers' assessments (one aspect of scientific quality) is not as strong as the relationship between peers' assessments and citations. Against the backdrop of the literature investigating the user population on the underlying platforms, this result was expectable (see, e.g., Yu, 2017). The platforms are not only used by scientists, but also by people who do not have the expertise to assess the quality of research. The results for the various altmetrics further show that Twitter, Wikipedia, Facebook, blogs, and news data are able to discriminate between the different quality levels (with statistical significance). This result might reflect that the faculty members do not only assess the quality of papers, but also other aspects which might be relevant for impact beyond science: suggesting new targets for drug discovery, challenging established dogma, or introducing a new practical/theoretical technique (see https://f1000.com/prime/about/whatis/how).

Since mentions in policy-related documents are not able to discriminate between different quality levels (some CIs partly overlap) as well as the other altmetrics, it seems that high-quality publications are not mentioned more frequently in policy-related documents than publications with lower quality. Another reason might be that the different sources which are tracked by Altmetric for this kind of altmetrics are not sufficient to reflect the whole picture of impact on policies. According to Haunschild and Bornmann (2017), more than 100 policy-related sources are currently tracked by Altmetric on December 19, 2015. Future studies should clarify whether the relationship of quality and mentions of papers in policy-related documents changes. Altmetric is adding more sources every month (see https://www.altmetric.com/about-our-data/our-sources/). A third reason for the non-significant result might be that the data are (still) too sparse for the use as altmetrics data source. Haunschild and Bornmann (2017) found with 0.5% a very low coverage of papers.

This study follows the important initiative of Thelwall (2017a, 2017b) to design new indicators for sparse data. Our study was the first independent attempt to investigate this indicator family empirically. The study focusses on a large publication set with a broad system of three quality levels (i.e., not mentioned by F1000, rather low $\overline{FFa}$ value, and rather high $\overline{FFa}$ value). Since our study demonstrates that the relationship of altmetrics and quality is not as strong as the relationship between citations and quality, it is interesting to see if altmetrics have a relationship with quality when finer quality levels are defined. Furthermore, it is unclear if our result can be transferred to scientific disciplines not rated by F1000. Thus, there is a high demand for further studies in this area. Since this family of indicators for sparse data is especially interesting for altmetrics data, we need further empirical studies.

# Acknowledgements

The bibliometric data used in this paper are from an in-house database developed and maintained by the Max Planck Digital Library (MPDL, Munich) and derived from the Science Citation Index Expanded (SCI-E), Social Sciences Citation Index (SSCI), and Arts and Humanities Citation Index (AHCI) prepared by Clarivate Analytics (formerly the IP and Science business of Thomson Reuters). The F1000Prime recommendations were taken from a data set retrieved from F1000 in November, 2017. The altmetrics data were taken from a data set retrieved from Altmetric on June 04, 2016 and stored in a local database and maintained by the Max Planck Institute for Solid State Research (Stuttgart). We would like to thank Mike Thelwall for helpful correspondence regarding calculation of the CIs for MNPC and EMNPC. We further thank two reviewers for their valuable recommendations to improve our manuscript.

# References

Adie, E. (2014). Taking the Alternative Mainstream. *Profesional De La Informacion, 23*(4), 349-351. doi: DOI 10.3145/epi.2014.jul.01.

Bik, H. M., & Goldstein, M. C. (2013). An Introduction to Social Media for Scientists. *PLoS Biol, 11*(4), e1001535. doi: 10.1371/journal.pbio.1001535.

Bollen, J., Van de Sompel, H., & Rodriguez, M. A. (2008). Towards usage-based impact metrics: first results from the MESUR project. Retrieved April 24, 2008, from http://arxiv.org/abs/0804.3791

Bornmann, L. (2011). Scientific peer review. *Annual Review of Information Science and Technology, 45*, 199-245.

Bornmann, L. (2014a). Do altmetrics point to the broader impact of research? An overview of benefits and disadvantages of altmetrics. *Journal of Informetrics, 8*(4), 895-903. doi: 10.1016/j.joi.2014.09.005.

Bornmann, L. (2014b). Validity of altmetrics data for measuring societal impact: A study using data from Altmetric and F1000Prime. *Journal of Informetrics, 8*(4), 935-950.

Bornmann, L. (2015a). Alternative metrics in scientometrics: A meta-analysis of research into three altmetrics. *Scientometrics, 103*(3), 1123-1144.

Bornmann, L. (2015b). Inter-rater reliability and convergent validity of F1000Prime peer review. *Journal of the Association for Information Science and Technology, 66*(12), 2415-2426.

Bornmann, L., & Haunschild, R. (2016a). How to normalize Twitter counts? A first attempt based on journals in the Twitter Index. *Scientometrics, 107*(3), 1405-1422. doi: 10.1007/s11192-016-1893-6.

Bornmann, L., & Haunschild, R. (2016b). Normalization of Mendeley reader impact on the reader- and paper-side: A comparison of the mean discipline normalized reader score (MDNRS) with the mean normalized reader score (MNRS) and bare reader counts. *Journal of Informetrics, 10*(3), 776-788.

Bornmann, L., Haunschild, R., & Marx, W. (2016). Policy documents as sources for measuring societal impact: How often is climate change research mentioned in policy-related documents? *Scientometrics, 109*(3), 1477–1495. doi: 10.1007/s11192-016-2115-y.

Butler, L., & McAllister, I. (2011). Evaluating university research performance using metrics. *European Political Science, 10*(1), 44-58. doi: 10.1057/eps.2010.13.

Claveau, F. (2016). There should not be any mystery: A comment on sampling issues in bibliometrics. *Journal of Informetrics, 10*(4), 1233-1240. doi: http://dx.doi.org/10.1016/j.joi.2016.09.009.

Colledge, L. (2014). *Snowball Metrics Recipe Book*. Amsterdam, the Netherlands: Snowball Metrics program partners.

Committee for Scientific and Technology Policy. (2014). *Assessing the impact of state interventions in research - techniques, issues and solutions*. Brussels, Belgium: Directorate for Science, Technology and Innovation.

Cumming, G. (2012). *Understanding the new statistics: effect sizes, confidence intervals, and meta-analysis*. London, UK: Routledge.

Cumming, G., & Finch, S. (2005). Inference by eye - Confidence intervals and how to read pictures of data. *American Psychologist, 60*(2), 170-180. doi: 10.1037/0003-066x.60.2.170.

Erdt, M., Nagarajan, A., Sin, S.-C. J., & Theng, Y.-L. (2016). Altmetrics: an analysis of the state-of-the-art in measuring research impact on social media. *Scientometrics*, 1-50. doi: 10.1007/s11192-016-2077-0.

Fairclough, R., & Thelwall, M. (2015). National research impact indicators from Mendeley readers. *Journal of Informetrics, 9*(4), 845-859. doi: http://dx.doi.org/10.1016/j.joi.2015.08.003.

Fleiss, J., Levin, B., & Paik, M. C. (2003). *Statistical methods for rates and proportions* (3. ed.). Hoboken, NJ, USA: Wiley.

Franceschet, M., & Costantini, A. (2011). The first Italian research assessment exercise: a bibliometric perspective. *Journal of Informetrics, 5*(2), 275-291. doi: DOI 10.1016/j.joi.2010.12.002.

Galloway, L. M., Pease, J. L., & Rauh, A. E. (2013). Introduction to Altmetrics for Science, Technology, Engineering, and Mathematics (STEM) Librarians. *Science & Technology Libraries, 32*(4), 335-345. doi: 10.1080/0194262X.2013.829762.

Garfield, E. (1979). *Citation indexing - its theory and application in science, technology, and humanities*. New York, NY, USA: John Wiley & Sons, Ltd.

Glänzel, W., & Schoepflin, U. (1995). A Bibliometric Study on Aging and Reception Processes of Scientific Literature. *Journal of Information Science, 21*(1), 37-53. doi: Doi 10.1177/016555159502100104.

Haunschild, R., & Bornmann, L. (2016). Normalization of Mendeley reader counts for impact assessment. *Journal of Informetrics, 10*(1), 62-73.

Haunschild, R., & Bornmann, L. (2017). How many scientific papers are mentioned in policy-related documents? An empirical investigation using Web of Science and Altmetric data. *Scientometrics, 110*(3), 1209-1216. doi: 10.1007/s11192-016-2237-2.

Haunschild, R., Schier, H., & Bornmann, L. (2016). Proposal of a minimum constraint for indicators based on means or averages. *Journal of Informetrics, 10*(2), 485-486. doi: 10.1016/j.joi.2016.03.003.

Haustein, S. (2016). Grand challenges in altmetrics: heterogeneity, data quality and dependencies. *Scientometrics, 108*(1), 413-423. doi: 10.1007/s11192-016-1910-9.

Haustein, S., Larivière, V., Thelwall, M., Amyot, D., & Peters, I. (2014). Tweets vs. Mendeley readers: How do these two social media metrics differ? *it – Information Technology 2014; 56(5): , 56*(5), 207-215.

Hollander, M., & Wolfe, D. A. (1999). *Nonparametric Statistical Methods*. New York, NY, USA: Wiley.

Kreiman, G., & Maunsell, J. H. R. (2011). Nine criteria for a measure of scientific output. *Frontiers in Computational Neuroscience, 5*(48). doi: 10.3389/fncom.2011.00048.

Li, X., Thelwall, M., & Giustini, D. (2012). Validating online reference managers for scholarly impact measurement. *Scientometrics, 91*(2), 461-471. doi: 10.1007/s11192-011-0580-x.

Mahdi, S., d'Este, P., & Neely, A. D. (2008). *Citation counts: are they good predictors of RAE scores? A bibliometric analysis of RAE 2001*. London, UK: Advanced Institute of Management Research.

Mantel, N., & Haenszel, W. (1959). Statistical Aspects of the Analysis of Data from Retrospective Studies of Disease. *Journal of the National Cancer Institute, 22*(4), 719-748.

Mas-Bleda, A., & Thelwall, M. (2016). Can alternative indicators overcome language biases in citation counts? A comparison of Spanish and UK research. *Scientometrics, 109*(3), 2007-2030. doi: 10.1007/s11192-016-2118-8.

McKay, S. (2012). Social policy excellence - peer review or metrics? Analyzing the 2008 Research Assessment Exercise in social work and social policy and administration. *Social Policy & Administration, 46*(5), 526-543. doi: 10.1111/j.1467-9515.2011.00824.x.

Mohammadi, E., Thelwall, M., & Kousha, K. (2016). Can Mendeley Bookmarks Reflect Readership? A Survey of User Motivations. *Journal of the Association for Information Science and Technology, 67*(5), 1198–1209.

National Information Standards Organization. (2016). *Outputs of the NISO Alternative Assessment Metrics Project*. Baltimore, MD, USA: National Information Standards Organization (NISO).

Neylon, C. (2014). Altmetrics: What are they good for? Retrieved October 6, 2014, from http://blogs.plos.org/opens/2014/10/03/altmetrics-what-are-they-good-for/#.VC8WETI0JAM.twitter

Plackett, R. L. (1974). *The analysis of categorical data*. London, UK: Chapman.

Priem, J. (2014). Altmetrics. In B. Cronin & C. R. Sugimoto (Eds.), *Beyond bibliometrics: harnessing multi-dimensional indicators of performance* (pp. 263-288). Cambridge, MA, USA: MIT Press.

Priem, J., Taraborelli, D., Groth, P., & Neylon, C. (2010). Altmetrics: a manifesto. Retrieved March 28, from http://altmetrics.org/manifesto/

Radhakrishna, S. (1965). Combination of results from several 2 x 2 contingency tables. *Biometrics, 21*, 86-98.

Rons, N. (2012). Partition-based Field Normalization: An approach to highly specialized publication records. *Journal of Informetrics, 6*(1), 1-10. doi: 10.1016/j.joi.2011.09.008.

Rons, N. (2014). Investigation of Partition Cells as a Structural Basis Suitable for Assessments of Individual Scientists. In P. Wouters (Ed.), *Proceedings of the science and technology indicators conference 2014 Leiden "Context Counts: Pathways to Master Big and Little Data"* (pp. 463-472). Leider, the Netherlands: University of Leiden.

Shema, H. (2014). Scholarly blogs are a promising altmetric source. *Research Trends*(37), 11-13.

Shema, H., Bar-Ilan, J., & Thelwall, M. (2012). Self-citation of bloggers in the science blogosphere. In A. Tokar, M. Beurskens, S. Keuneke, M. Mahrt, I. Peters, C. Puschmann, K. Weller & T. van Treeck (Eds.), *Science and the Internet* (pp. 183-192). Düsseldorf, Germany: Düsseldorf University Press.

Sheskin, D. (2007). *Handbook of parametric and nonparametric statistical procedures* (4th ed.). Boca Raton, FL, USA: Chapman & Hall/CRC.

Smith, A., & Eysenck, M. (2002). *The correlation between RAE ratings and citation counts in psychology*. London: Department of Psychology, Royal Holloway, University of London, UK.

Sugimoto, C. (2016). "Attention is not impact" and other challenges for altmetrics. Retrieved September, 9, 2016, from https://hub.wiley.com/community/exchanges/discover/blog/2015/06/23/attention-is-not-impact-and-other-challenges-for-altmetrics

Taylor, M. (2013). Towards a common model of citation: some thoughts on merging altmetrics and bibliometrics. *Research Trends*(35), 19-22.

Thelwall, M. (2016). Three practical field normalised alternative indicator formulae for research evaluation. Retrieved from https://arxiv.org/abs/1612.01431

Thelwall, M. (2017a). Three practical field normalised alternative indicator formulae for research evaluation. *Journal of Informetrics, 11*(1), 128-151. doi: 10.1016/j.joi.2016.12.002.

Thelwall, M. (2017b). *Web Indicators for Research Evaluation: A Practical Guide*. London, UK: Morgan & Claypool.

Thelwall, M., & Kousha, K. (2015a). Web Indicators for Research Evaluation. Part 1: Citations and Links to Academic Articles from the Web. *Profesional De La Informacion, 24*(5), 587-606. doi: 10.3145/epi.2015.sep.08.

Thelwall, M., & Kousha, K. (2015b). Web Indicators for Research Evaluation. Part 2: Social Media Metrics. *Profesional De La Informacion, 24*(5), 607-620. doi: 10.3145/epi.2015.sep.09.

Thelwall, M., Kousha, K., Dinsmore, A., & Dolby, K. (2016). Alternative metric indicators for funding scheme evaluations. *Aslib Journal of Information Management, 68*(1), 2-18. doi: doi:10.1108/AJIM-09-2015-0146.

Waltman, L., & Costas, R. (2014). F1000 Recommendations as a Potential New Data Source for Research Evaluation: A Comparison With Citations. *Journal of the Association for Information Science and Technology, 65*(3), 433-445. doi: 10.1002/asi.23040.

Waltman, L., van Eck, N. J., van Leeuwen, T. N., Visser, M. S., & van Raan, A. F. J. (2011). Towards a new crown indicator: an empirical analysis. *Scientometrics, 87*(3), 467-481. doi: 10.1007/s11192-011-0354-5.

Williams, R., & Bornmann, L. (2016). Sampling issues in bibliometric analysis. *Journal of Informetrics, 10*(4), 1253-1257.

Wilsdon, J., Allen, L., Belfiore, E., Campbell, P., Curry, S., Hill, S., . . . Johnson, B. (2015). *The Metric Tide: Report of the Independent Review of the Role of Metrics in Research Assessment and Management*. Bristol, UK: Higher Education Funding Council for England (HEFCE).

Work, S., Haustein, S., Bowman, T. D., & Larivière, V. (2015). *Social Media in Scholarly Communication. A Review of the Literature and Empirical Analysis of Twitter Use by SSHRC Doctoral Award Recipients*. Montreal, Canada: Canada Research Chair on the Transformations of Scholarly Communication, University of Montreal.

Wouters, P., & Costas, R. (2012). *Users, narcissism and control – tracking the impact of scholarly publications in the 21st century*. Utrecht, The Netherlands: SURFfoundation.

Wouters, P., Thelwall, M., Kousha, K., Waltman, L., de Rijcke, S., Rushforth, A., & Franssen, T. (2015). *The Metric Tide: Correlation analysis of REF2014 scores and metrics (Supplementary Report II to the Independent Review of the Role of Metrics in Research Assessment and Management)*. London, UK: Higher Education Funding Council for England (HEFCE).

Yu, H. (2017). Context of altmetrics data matters: an investigation of count type and user category. *Scientometrics, 111*(1), 267-283. doi: 10.1007/s11192-017-2251-z.

Zahedi, Z., Costas, R., & Wouters, P. (2014). How well developed are altmetrics? A cross-disciplinary analysis of the presence of 'alternative metrics' in scientific publications. *Scientometrics, 101*(2), 1491-1513. doi: 10.1007/s11192-014-1264-0.